\documentclass[preprint,aps,amsmath,superscriptaddress,nofootinbib,tightenlines,showpacs]{revtex4}
\usepackage{amsmath}
\usepackage{amsfonts}
\usepackage{amssymb}
\usepackage{latexsym}
\usepackage{graphicx}
\usepackage{bm}
\usepackage{epsfig}
\usepackage[english]{babel}

\begin{document}

%\preprint{\vbox{ \hbox{}   \hbox{} }}

\title{Modified entropic gravity revisited}
\author{Tower Wang\footnote{Electronic address: twang@phy.ecnu.edu.cn}}
\affiliation{Department of Physics, East China Normal University,\\
Shanghai 200241, China\\ \vspace{0.2cm}}
\date{\today\\ \vspace{1cm}}
\begin{abstract}
Inspired by Verlinde's idea, some modified versions of entropic
gravity have appeared in the literature. Extending them in a unified
formalism, we derive the generalized gravitational equations
accordingly. From gravitational equations, the energy-momentum
conservation law and cosmological equations are investigated. The
covariant conservation law of energy-momentum tensor severely
constrains viable modifications of entropic gravity. A discrepancy
arises when two independent methods are applied to the homogeneous
isotropic universe, posing a serious challenge to modified models of
entropic gravity.
\end{abstract}


\maketitle

%\tighten

%%%%%%%%%%%%%%%%%%%%%%%%%%%%%%%%%%%%%%%%%%

%\tableofcontents
\section{Introduction}\label{sect-intro}
In the past forty years, our understanding about the nature of
gravity has been much enriched by the thermodynamics of gravity. The
black hole thermodynamics was well established, and the holographic
principle was rigorously realized in AdS/CFT correspondence.
Encouraged by these achievements, it was conjectured that gravity is
an emergent phenomenon. In other words, gravity may not be a
fundamental force.

An intriguing way towards emergent gravity is making Einstein's
equations from the Clausius relation between entropy change and heat
flux \cite{Jacobson:1995ab}, or more recently, from the
equipartition relation between energy and bits
\cite{Verlinde:2010hp}. In the latter scenario, gravity is deemed as
an entropic force, and the derivation of gravitational equations
\cite{Verlinde:2010hp} relies on three ingredients: the Unruh law
\cite{Unruh:1976db} identifying temperature with the local
acceleration; the holographic principle ensuring that the number of
bits is proportional to the area of holographic screen; and an
equipartition rule \cite{Padmanabhan:2009kr} relating energy or mass
to the temperature and the number of bits.

Soon after the proposal, entropic gravity was widely
studied.\footnote{The readers may refer to
\cite{Zhao:2010qw,Tian:2010uy,Liu:2010na,KowalskiGlikman:2010ms,Myung:2010jv,Konoplya:2010ak,Gao:2010yy,Cai:2010zw}
as a partial list and references therein.} Interestingly, a lot of
modified models of entropic gravity emerged. Almost all of them can
be classified broadly into three categories, corresponding to
modifying one of the ingredients above. First, applying a different
relation between temperature and acceleration, modified Newtonian
dynamics was reinterpreted in
\cite{Ho:2010ca,Pikhitsa:2010nd,Klinkhamer:2011un} through the
entropic force. Second, deformed entropy-area relations motivated
further investigations on modified gravity in
\cite{Zhang:2010hi,Wei:2010wwa,Modesto:2010rm,Myung:2010kd,Sheykhi:2010wm,Pan:2010eu,Setare:2010ct,Liu:2010zn,Sheykhi:2010yq,Hendi:2010xr,ChangYoung:2011rb}.
Third, Refs.
\cite{Gao:2010fw,Li:2010si,Wei:2010am,Chang:2010be,Kiselev:2010xi,Neto:2010ds,Abreu:2012ax,SheykhiSarab,Wang:2012pk}
explored the possibility of entropic gravity with more complicated
equipartition relations. Exceptions include changing more than one
ingredients \cite{Liu:2010wp} or considering corrections from the
uncertainty principle
\cite{Vancea:2010vf,Ghosh:2010hz,ShalytMargolin:2012zh}.

In Verlinde's enrtopic gravity scenario, the Unruh law $k_BT=\hbar
a/(2\pi c)$, the holographic relation $N=Ac^3/(G\hbar)$ and the
equipartition rule $E=Nk_BT/2$ are combined to give
\begin{equation}\label{EaA}
E=\frac{aAc^2}{4\pi G}.
\end{equation}
Taking $E$ as the total energy inside a closed holographic surface,
and identifying $A$ as the surface area and $a$ the (red-shifted)
surface acceleration, one is able to build up the Newton's
(Einstein's) gravitational equations \cite{Verlinde:2010hp}.

It is not impossible that Einstein gravity holds exactly at all
energy scales, no matter how strong or weak the gravitational field
is. There is also possibility that gravity does not have an entropic
origin. Otherwise, if gravity is really an entropic force and gets
modified in some regions, we may have to take modified entropic
gravity seriously.

In the present paper, we start our adventure by observing that most
modified entropic gravity models summarized above, whichever
ingredient they modify, can be formally expressed by inserting a
factor $f(a,A)$ into Eq. \eqref{EaA},
\begin{equation}\label{EaAf}
E=\frac{aAc^2}{4\pi G}f(a,A).
\end{equation}
Here $f(a,A)$ is a model-dependent function of acceleration and area
of the holographic surface. In the relativistic case, acceleration
$a$ should be red-shifted like temperature.

The rest of our paper is organized as follows. Starting from
relation \eqref{EaAf}, we will derive gravitational equations in
Sec. \ref{subsect-rel} and deal with the static weak field limit in
Sec. \ref{subsect-weak}. Although the gravitational equations look
fine, we find in Sec. \ref{sect-cons} that viable models of modified
entropic gravity are constrained tightly by the covariant
conservation of energy-momentum. Under the assumption of
Friedmann-Lema\^{i}tre-Robertson-Walker (FLRW) metric, in Sec.
\ref{sect-cosmo} we will write down cosmological equations, which
further constrain the modified entropic gravity models as viable
theories to understand our Universe. In Sec. \ref{sect-disc} we
discuss possible implications of our results. Some useful formulae
for FLRW spacetime are relegated to Appendix \ref{app-form}.
Throughout this paper, we will follow the convention of notations in
\cite{Wang:2012pk}.

\section{Gravitational equations}\label{sect-grav}
\subsection{Field equations}\label{subsect-rel}
To get the relativistic gravitational field equations, parallel to
Ref. \cite{Verlinde:2010hp} we begin with a static background which
has a time-like Killing vector $\xi^{\mu}$ normalized as
\begin{equation}\label{Knorm}
\xi_{\mu}\xi^{\mu}=-e^{2\phi}
\end{equation}
with the Newtonian potential $\phi$. Introducing a vector
$\mathcal{N}^{\mu}$ outward normal to the holographic surface
$\mathcal{S}$, we can express the red-shifted acceleration as
\begin{equation}\label{acc}
a=e^{\phi}\mathcal{N}^{\mu}\nabla_{\mu}\phi.
\end{equation}
Then by virtue of $E=Mc^2$ it is easy to write Eq. \eqref{EaAf} in
the integral form
\begin{equation}\label{amassA}
M=\frac{1}{4\pi G}\int_{\mathcal{S}}fe^{\phi}\nabla\phi\cdot dA,
\end{equation}
where $dA_{\mu}=|dA|\mathcal{N}_{\mu}$ and
\begin{equation}\label{dAv}
|dA|=dx^{\alpha}dx^{\beta}\epsilon_{\kappa\lambda\alpha\beta}e^{-\phi}\xi^{\kappa}\mathcal{N}^{\lambda}.
\end{equation}

In this subsection, we will derive the modified field equations
following the convention of notations in \cite{Wang:2012pk}. To
avoid repetition, some results will be quoted directly from
\cite{Wang:2012pk}. Although the present subsection is clear in
outline and new details, for old details of the quoted results, we
refer the reader to Sec. IVA and Appendix A of \cite{Wang:2012pk}.

Here are some essential points from Ref. \cite{Wang:2012pk}. In the
appendix of \cite{Wang:2012pk}, it has been demonstrated that for a
general function $f$, e.g. $f(a,A)$, the expression of mass
\eqref{amassA} can be rearranged as
\begin{equation}\label{Jacobson4}
M=\frac{1}{4\pi G}\int_{\mathcal{V}}[f\mathcal{R}_{~\sigma}^{\rho}\xi^{\sigma}n_{\rho}+(\nabla_{\sigma}f)(\nabla^{\rho}\xi^{\sigma})n_{\rho}]dV,
\end{equation}
in which $\mathcal{V}$ is the 3-dimensional volume bounded by the
holographic screen $\mathcal{S}$, $n^{\mu}$ is a future-directed
vector normal to $\mathcal{V}$, and $\mathcal{R}_{\mu\nu}$ denotes
the Ricci tensor. As also mentioned in the appendix of
\cite{Wang:2012pk}, one may recast the second term of
\eqref{Jacobson4} into
\begin{equation}\label{2term}
(\nabla_{\sigma}f)(\nabla^{\rho}\xi^{\sigma})n_{\rho}=n_{\rho}\nabla^{\rho}(\xi^{\sigma}\nabla_{\sigma}f)-n_{\rho}\xi^{\sigma}\nabla^{\rho}\nabla_{\sigma}f.
\end{equation}
Moreover, we have shown in Ref. \cite{Wang:2012pk} that if one can
prove $\xi^{\sigma}\nabla_{\sigma}f=0$, then gravitational equations
are of the form
\begin{equation}\label{Einstein}
f\mathcal{R}_{\mu\nu}-\nabla_{\mu}\nabla_{\nu}f=8\pi G\left(\mathcal{T}_{\mu\nu}-\frac{1}{2}g_{\mu\nu}\mathcal{T}\right),
\end{equation}
where $\mathcal{T}_{\mu\nu}$ is the energy-momentum tensor. In
Appendix A of \cite{Wang:2012pk}, it was established that
$\xi^{\sigma}\nabla_{\sigma}(e^{\phi}\mathcal{N}^{\mu}\nabla_{\mu}\phi)=0$,
namely $\xi^{\sigma}\nabla_{\sigma}a=0$, hence
$\xi^{\sigma}\nabla_{\sigma}f(a)=0$ and equations \eqref{Einstein}
are correct if $f$ is a function purely of acceleration \eqref{acc}.

In the rest of this subsection, we will show that
$\xi^{\sigma}\nabla_{\sigma}f(a,A)=0$ and thus \eqref{Einstein}
continue to be correct when $f$ is a function of both acceleration
$a$ and area $A$ generally. This can be achieved by proving the
equality
\begin{equation}\label{xidA}
\xi^{\sigma}\nabla_{\sigma}A=0.
\end{equation}
To do this, we note
\begin{eqnarray}\label{xidA1}
\nonumber \xi^{\sigma}\nabla_{\sigma}A&=&\xi^{\sigma}\nabla_{\sigma}\int_{\mathcal{S}}dx^{\alpha}dx^{\beta}\epsilon_{\mu\nu\alpha\beta}e^{-\phi}\xi^{\mu}\mathcal{N}^{\nu}\\
&=&\int_{\mathcal{S}}dx^{\alpha}dx^{\beta}\xi^{\sigma}\nabla_{\sigma}\left(\epsilon_{\mu\nu\alpha\beta}e^{-\phi}\xi^{\mu}\mathcal{N}^{\nu}\right).
\end{eqnarray}
In the second line, we have used normalization condition
\eqref{Knorm} and the fact that the holographic screen $\mathcal{S}$
corresponds to the equipotential surface \cite{Verlinde:2010hp}.

To proceed, we recall that the holographic screen $\mathcal{S}$ is a
2-dimensional surface. Restricted to this surface, we can define a
2-dimensional tensor
$\omega^{\alpha\beta}=\xi_{\kappa}\mathcal{N}_{\lambda}\epsilon^{\kappa\lambda\alpha\beta}$.
Making use of the equality
\begin{equation}\label{contract2}
\epsilon^{\kappa\lambda\alpha\beta}\epsilon_{\mu\nu\alpha\beta}=-4\delta^{\kappa}_{~[\mu}\delta^{\lambda}_{~\nu]}=\frac{-4}{2!}(\delta^{\kappa}_{~\mu}\delta^{\lambda}_{~\nu}-\delta^{\kappa}_{~\nu}\delta^{\lambda}_{~\mu}),
\end{equation}
it is not hard to see
\begin{equation}\label{xidA2}
\omega^{\alpha\beta}\xi^{\sigma}\nabla_{\sigma}\left(\epsilon_{\mu\nu\alpha\beta}e^{-\phi}\xi^{\mu}\mathcal{N}^{\nu}\right)=e^{-\phi}\xi^{\mu}\mathcal{N}^{\nu}\xi_{\kappa}N_{\lambda}\epsilon^{\kappa\lambda\alpha\beta}\xi^{\sigma}\nabla_{\sigma}\epsilon_{\mu\nu\alpha\beta}-2(\xi_{\mu}N_{\nu}-\xi_{\nu}N_{\mu})\xi^{\sigma}\nabla_{\sigma}\left(e^{-\phi}\xi^{\mu}\mathcal{N}^{\nu}\right).
\end{equation}
It is interesting to manipulate indices of the first term of
\eqref{xidA2} as
\begin{eqnarray}\label{xidA3}
\nonumber e^{-\phi}\xi^{\mu}\mathcal{N}^{\nu}\xi_{\kappa}N_{\lambda}\epsilon^{\kappa\lambda\alpha\beta}\xi^{\sigma}\nabla_{\sigma}\epsilon_{\mu\nu\alpha\beta}&=&e^{-\phi}\xi_{\mu}\mathcal{N}_{\nu}\xi^{\kappa}N^{\lambda}\epsilon_{\kappa\lambda\alpha\beta}\xi^{\sigma}\nabla_{\sigma}\epsilon^{\mu\nu\alpha\beta}\\
\nonumber &=&e^{-\phi}\xi_{\kappa}\mathcal{N}_{\lambda}\xi^{\mu}N^{\nu}\epsilon_{\mu\nu\alpha\beta}\xi^{\sigma}\nabla_{\sigma}\epsilon^{\kappa\lambda\alpha\beta}\\
\nonumber &=&e^{-\phi}\xi^{\mu}\mathcal{N}^{\nu}\xi_{\kappa}N_{\lambda}\xi^{\sigma}\nabla_{\sigma}\left(\frac{1}{2}\epsilon^{\kappa\lambda\alpha\beta}\epsilon_{\mu\nu\alpha\beta}\right)\\
%\nonumber &=&e^{-\phi}\xi^{\mu}\mathcal{N}^{\nu}\xi_{\kappa}N_{\lambda}\xi^{\sigma}\nabla_{\sigma}\left(\delta^{\kappa}_{~\nu}\delta^{\lambda}_{~\mu}-\delta^{\kappa}_{~\mu}\delta^{\lambda}_{~\nu}\right)\\
&=&0.
\end{eqnarray}
The last line follows from \eqref{contract2} and
$\nabla_{\sigma}\delta^{\mu}_{~\nu}=0$. In Ref. \cite{Wang:2012pk}
it has been shown that
\begin{equation}
\xi^{\mu}\nabla_{\mu}\phi=-e^{-2\phi}\xi^{\mu}\xi^{\nu}\nabla_{\mu}\xi_{\nu}=0,~~~\mathcal{N}_{\nu}\nabla^{\mu}\mathcal{N}^{\nu}=0.
\end{equation}
Therefore the second term of \eqref{xidA2} is simplified,
\begin{equation}\label{xidA4}
-2(\xi_{\mu}N_{\nu}-\xi_{\nu}N_{\mu})\xi^{\sigma}\nabla_{\sigma}\left(e^{-\phi}\xi^{\mu}\mathcal{N}^{\nu}\right)=2e^{-\phi}\xi_{\nu}N_{\mu}\xi^{\sigma}\nabla_{\sigma}\left(\xi^{\mu}\mathcal{N}^{\nu}\right)=0.
\end{equation}
Here the orthogonal relation $\mathcal{N}^{\mu}\xi_{\mu}=0$ is taken
into account. Putting Eqs. \eqref{xidA2}, \eqref{xidA3} and
\eqref{xidA4} together, it is apparent that
\begin{equation}\label{xidA5}
\omega^{\alpha\beta}\xi^{\sigma}\nabla_{\sigma}\left(\epsilon_{\mu\nu\alpha\beta}e^{-\phi}\xi^{\mu}\mathcal{N}^{\nu}\right)=0.
\end{equation}
At the same time, restricted to the holographic screen
$\mathcal{S}$, all 2-dimensional anti-symmetric covariant tensor
should be proportional to $\omega_{\alpha\beta}$ and thus
proportional to each other in the same basis. That is to say,
\begin{equation}\label{xidA6}
\left.\xi^{\sigma}\nabla_{\sigma}\left(\epsilon_{\mu\nu\alpha\beta}e^{-\phi}\xi^{\mu}\mathcal{N}^{\nu}\right)\right|_{\mathcal{S}}=0
\end{equation}
and subsequently
\begin{equation}\label{xidA7}
\int_{\mathcal{S}}dx^{\alpha}dx^{\beta}\xi^{\sigma}\nabla_{\sigma}\left(\epsilon_{\mu\nu\alpha\beta}e^{-\phi}\xi^{\mu}\mathcal{N}^{\nu}\right)=0.
\end{equation}
This conclude our proof of Eq. \eqref{xidA}.

Since we have demonstrated $\xi^{\sigma}\nabla_{\sigma}A=0$ in the
above, and Ref. \cite{Wang:2012pk} has verified that
$\xi^{\sigma}\nabla_{\sigma}a=0$, it is straightforward to obtain
\begin{equation}
\xi^{\sigma}\nabla_{\sigma}f(a,A)=f_{,a}\xi^{\sigma}\nabla_{\sigma}a+f_{,A}\xi^{\sigma}\nabla_{\sigma}A=0
\end{equation}
as promised earlier in this section. We can conclude that, for
modified entropic gravity models of form \eqref{EaAf}, when $f$ is a
general function of acceleration $a$ and area $A$, the gravitational
field equations are given by \eqref{Einstein}. In Verlinde's
entropic gravity model, $f=1$, and Eqs. \eqref{Einstein} reduce to
the Einstein equations.

\subsection{Static weak field limit}\label{subsect-weak}
The static weak field limit provides not only a crosscheck of our
calculations in Sec. \ref{subsect-rel}, but also a playground to
test modified models of entropic gravity. Hence it would be valuable
to take a closer look at this limit.

First, we note that the Newtonian limit corresponds to the
00-component of relativistic gravitational equations. For this
component, the second term of the left hand side of \eqref{Einstein}
reads
\begin{equation}
-\nabla_{0}\nabla_{0}f=-\partial_{0}\partial_{0}f+\Gamma^{\mu}_{00}\partial_{\mu}f.
\end{equation}
In the static case, both temperature and entropy of the holographic
surface is time-independent, hence $\partial_{0}\partial_{0}f$
vanishes. The weak gravitational field $g_{\mu\nu}$ can be expanded
near the Minkowski metric $\eta_{\mu\nu}$ as
\begin{equation}
g_{\mu\nu}=\eta_{\mu\nu}+h_{\mu\nu},~~~|h_{\mu\nu}|\ll1.
\end{equation}
For a static metric, we have $\partial_{0}g_{\mu\nu}=0$ and
\begin{equation}
\Gamma^{\mu}_{00}=-\frac{1}{2}\eta^{\mu\lambda}\partial_{\lambda}h_{00},~~~\mathcal{R}_{00}=-\frac{1}{2}\nabla^{\lambda}\nabla_{\lambda}h_{00}
\end{equation}
in which $h_{00}\simeq-2\phi$. Therefore the 00-component of
\eqref{Einstein} on the left hand side becomes
\begin{eqnarray}
\nonumber f\mathcal{R}_{00}-\nabla_{0}\nabla_{0}f&\simeq&f\eta^{\mu\lambda}\partial_{\mu}\partial_{\lambda}\phi+\eta^{\mu\lambda}(\partial_{\lambda}\phi)\partial_{\mu}f\\
&=&\eta^{\mu\lambda}\partial_{\mu}(f\partial_{\lambda}\phi).
\end{eqnarray}

On the other hand, in the Newtonian limit, the stress-energy tensor
$\mathcal{T}_{\mu\nu}=\rho u_{\mu}u_{\nu}$ with
$u_{\mu}=(\sqrt{-g_{00}},0,0,0)$. As a result, we find in this limit
the Poisson's equation is generalized to
\begin{equation}\label{Gaussd}
\delta^{ij}\partial_{i}(f\partial_{j}\phi)\simeq4\pi G\rho
\end{equation}
where indices $i$, $j$ represent spatial coordinates. In the case
$f=1$, this equation reduces to the Newton's law of gravity.

The above generalized Poisson's equation can be also obtained
directly by writing \eqref{amassA} as
\begin{equation}
\int_{\mathcal{V}}\rho dV\simeq\frac{1}{4\pi G}\int_{\mathcal{S}}\delta^{ij}f\partial_{j}\phi\cdot dA_{i}
\end{equation}
in the static weak field limit and applying the divergence theorem.
This confirms that our calculation is consistent.

A brief comment is in order here. If function $f$ deviates obviously
from constant or from 1, with the help of Eq. \eqref{Gaussd}, we can
test modified entropic gravity models against observations such as
planetary orbits.

\section{Energy-momentum conservation law}\label{sect-cons}
The conservation of energy is a fundamental law in physics. In
general relativity, it has a nice generalization: the covariant
conservation of energy-momentum tensor, formulated as
$\nabla_{\nu}\mathcal{T}_{\mu}^{~\nu}=0$ and guaranteed by the fact
that the covariant divergence of Einstein tensor is zero,
\begin{equation}
\nabla_{\nu}\left(\mathcal{R}_{\mu}^{~\nu}-\frac{1}{2}\delta_{\mu}^{~\nu}\mathcal{R}\right)=0.
\end{equation}
In modified entropic gravity, gravitational field equations
\eqref{Einstein}  can be transformed to
%\begin{equation}
%-f\mathcal{R}+\nabla^2f=8\pi G\mathcal{T}.
%\end{equation}
\begin{equation}\label{Einstein1}
f\left(\mathcal{R}_{\mu\nu}-\frac{1}{2}g_{\mu\nu}\mathcal{R}\right)-\left(\nabla_{\mu}\nabla_{\nu}f-\frac{1}{2}g_{\mu\nu}\nabla^2f\right)=8\pi G\mathcal{T}_{\mu\nu},
\end{equation}
more complicated than the counterparts in Einstein gravity, so in
this new situation we have to reconsider the law of energy-momentum
conservation.

For this purpose, we will work out the covariant derivative of the
left hand side of Eq. \eqref{Einstein1}. Remind that for a vector
$\nabla^{\nu}f$, one has
%\begin{equation}\label{Rform}
%\nabla_{\sigma}\nabla_{\mu}v^{\nu}-\nabla_{\mu}\nabla_{\sigma}v^{\nu}=\mathcal{R}_{\sigma\mu~\lambda}^{~~~\nu}v^{\lambda}
%\end{equation}
\begin{equation}
\nabla_{\nu}\nabla_{\mu}\nabla^{\nu}f-\nabla_{\mu}\nabla_{\nu}\nabla^{\nu}f=\mathcal{R}_{\mu\lambda}\nabla^{\lambda}f
\end{equation}
by definition of the Riemann tensor. Thanks to this identity, after
a little algebra, one may directly demonstrate
%\begin{eqnarray}
%\nonumber 8\pi G\nabla_{\nu}\mathcal{T}_{\mu}^{~\nu}&=&f\nabla_{\nu}\left(\mathcal{R}_{\mu}^{~\nu}-\frac{1}{2}\delta_{\mu}^{~\nu}\mathcal{R}\right)+\left(\mathcal{R}_{\mu}^{~\nu}-\frac{1}{2}\delta_{\mu}^{~\nu}\mathcal{R}\right)\nabla_{\nu}f-\nabla_{\nu}\nabla_{\mu}\nabla^{\nu}f+\frac{1}{2}\nabla_{\mu}\nabla^2f\\
%\nonumber &=&\left(\mathcal{R}_{\mu}^{~\nu}-\frac{1}{2}\delta_{\mu}^{~\nu}\mathcal{R}\right)\nabla_{\nu}f-\nabla_{\nu}\nabla_{\mu}\nabla^{\nu}f+\frac{1}{2}\nabla_{\mu}\nabla^2f\\
%\nonumber &=&-\frac{1}{2}\mathcal{R}\nabla_{\mu}f-\frac{1}{2}\nabla_{\mu}\nabla^2f\\
%&=&-\frac{1}{2}\left(\mathcal{R}\nabla_{\mu}f+\nabla_{\mu}\nabla^2f\right).
%\end{eqnarray}
\begin{equation}
8\pi G\nabla_{\nu}\mathcal{T}_{\mu}^{~\nu}=-\frac{1}{2}\left(\mathcal{R}\nabla_{\mu}f+\nabla_{\mu}\nabla^2f\right).
\end{equation}
That is, the energy-momentum tensor cannot be covariantly conserved
in modified entropic gravity unless
\begin{equation}\label{emcons}
\mathcal{R}\nabla_{\mu}f+\nabla_{\mu}\nabla^2f=0.
\end{equation}
Only a narrow subset of models are permitted by this requirement.

The situation is reminiscent of the so-called Chern-Simons modified
general relativity \cite{Jackiw:2003pm}, in which the
energy-momentum conservation is unwarranted by gravitational
equations. In the literature, this is not taken as a fatal defect of
the Chern-Simons modified gravity, though one should be careful with
it when seeking for exact metric solutions. Likewise, we can say Eq.
\eqref{emcons} is a criterion of admissible metrics in modified
models of entropic gravity.

\section{Cosmological equations}\label{sect-cosmo}
In Refs. \cite{Shu:2010nv,Cai:2010hk}, Friedmann equations are
obtained from Verlinde's model \cite{Verlinde:2010hp}. In modified
models \eqref{EaAf} of entropic gravity, it is possible to derive
cosmological equations along the same line. To avoid confusions in
notation, we use $R(t)$ to denote the scale factor. Another method
to write down cosmological equations is by directly applying
gravitational equations \eqref{Einstein} to cosmological background.
We will explore both approaches, assuming the FLRW metric
\begin{equation}\label{FLRW}
ds^2=-dt^2+R^2\left[dr^2+r^2(d\theta^2+\sin^2\theta d\varphi^2)\right].
\end{equation}

Let us consider Eq. \eqref{EaAf} on a sphere with physical radius
$Rr$ in the FLRW universe. On the one hand, the red-shifted
acceleration and area are now expressed as $a=-\ddot{R}r$ and
$A=4\pi R^2r^2$. On the other hand, $E=Mc^2$, and the active
gravitational mass \cite{Padmanabhan:2003pk} is given by
%\begin{equation}
%M=\int_{\mathcal{V}}(\rho+3p)dV=\int_{\mathcal{V}}(\rho+3p)4\pi R^3r^2dr=\frac{4}{3}\pi R^3r^3(\rho+3p)
%\end{equation}
\begin{equation}
M=\int_{\mathcal{V}}(\rho+3p)dV=\int_{\mathcal{V}}(\rho+3p)4\pi R^3r^2dr
\end{equation}
for a perfect fluid whose stress-energy tensor
$\mathcal{T}_{\mu\nu}=(\rho+p)u_{\mu}u_{\nu}+pg_{\mu\nu}$ with
$u_{\mu}u^{\mu}=-1$. In accordance with metric \eqref{FLRW}, it is
convenient to choose $u_{\mu}=(1,0,0,0)$. Then Eq. \eqref{EaAf}
leads to the acceleration equation
\begin{equation}\label{Friedmann1}
\frac{\ddot{R}}{R}f=-\frac{4\pi G}{r^3}\int_{\mathcal{V}}(\rho+3p) r^2dr.
\end{equation}
This equation combined with continuity equation
\begin{equation}\label{contin}
\dot{\rho}+\frac{3\dot{R}}{R}(\rho+p)=0
\end{equation}
dictates the background dynamics of cosmology. Especially, putting
them together, we find
\begin{equation}\label{Friedmann2}
\dot{R}\ddot{R}f=\frac{4\pi G}{r^3}\frac{d}{dt}\int_{\mathcal{V}}\rho R^2r^2dr.
\end{equation}

It is useful to recall that continuity condition \eqref{contin}
comes from the covariant conservation law of energy-momentum.
Besides this condition, energy-momentum conservation
$\nabla_{\nu}\mathcal{T}_{\mu}^{~\nu}=0$ indicates also
$\partial_{r}p=\partial_{\theta}p=\partial_{\varphi}p=0$. With these
conditions imposed, Eqs. \eqref{Friedmann1} and \eqref{Friedmann2}
are integrated as
\begin{equation}
\frac{4\pi G}{r^3}\int_{\mathcal{V}}\rho R^2r^2dr=-R\ddot{R}f-4\pi GpR^2=\int\dot{R}\ddot{R}fdt,
\end{equation}
yielding a differential equation
\begin{equation}
\partial_{t}(R^2\ddot{R}\partial_{r}f)=0
\end{equation}
which is apparently solved by
\begin{equation}
f=\frac{g_1(r)}{R^2\ddot{R}}+g_2(t).
\end{equation}
Remember that $f$ is a function of $a$ and $A$, so the expression of
$f$ should have the form
\begin{equation}\label{f1}
f=\frac{C_1}{aA}+g_2\left(aA^{-1/2}\right).
\end{equation}
Here $C_1$ is a constant, while $g_2$ is a function of
$\ddot{R}R^{-1}$.

As implied by Eq. \eqref{Friedmann1}, if $C_1\neq0$, then density
$\rho$ will be dependent of radial coordinate $r$, violating the
Copernican principle. Therefore, to get rid of exotic cosmology with
radial dependence, one may set $C_1=0$ and treat $f$ as a function
of $aA^{-1/2}$. It is remarkable that the cosmologically viable
modified entropic gravity is restricted to such a small subset of
models, simply by covariant conservation of energy momentum and the
Copernican principle. We will investigate this subset of
cosmologically viable models elsewhere.

%\begin{eqnarray}
%\nonumber \mathcal{R}_{\mu\nu}&=&\left(-\frac{3\ddot{R}}{R},g_{ij}\left(\frac{\ddot{R}}{R}+\frac{2\dot{R}^2}{R^2}\right)\right),\\
%8\pi G\left(\mathcal{T}_{\mu\nu}-\frac{1}{2}g_{\mu\nu}\mathcal{T}\right)&=&4\pi G\left(\rho+3p,g_{ij}(\rho-p)\right).
%\end{eqnarray}
Since we have established gravitational equations \eqref{Einstein}
in Sec. \ref{subsect-rel}, there is another approach to cosmological
equations. That is applying Eqs. \eqref{Einstein} to the FLRW metric
\eqref{FLRW}. By doing this, we find
\begin{eqnarray}\label{Einstein-FLRW}
\nonumber -\frac{3\ddot{R}}{R}f-\partial_{t}^2f&=&4\pi G(\rho+3p),\\
\nonumber \partial_{t}\partial_{r}f-\frac{\dot{R}}{R}\partial_{r}f&=&0,\\
\nonumber \left(\frac{\ddot{R}}{R}+\frac{2\dot{R}^2}{R^2}\right)f+\frac{\dot{R}}{R}\partial_{t}f-\frac{1}{R^2}\partial_{r}^2f&=&4\pi G(\rho-p),\\
\left(\frac{\ddot{R}}{R}+\frac{2\dot{R}^2}{R^2}\right)f+\frac{\dot{R}}{R}\partial_{t}f-\frac{1}{R^2r}\partial_{r}f&=&4\pi G(\rho-p).
\end{eqnarray}
These gravitational equations put very stringent limits on $f$ by
differential equations
\begin{equation}
\partial_{r}\left(\frac{1}{r}\partial_{r}f\right)=\partial_{t}\left(\frac{1}{R}\partial_{r}f\right)=0,
\end{equation}
whose solution is
\begin{equation}
f=C_1Rr^2+g_2(t).
\end{equation}
Because $f$ is a function of $a$ and $A$, the admissible expression
should be of the form
\begin{equation}\label{f2}
f=g_2\left(aA^{-1/2}\right).
\end{equation}
Here $g_2$ is a function of $\ddot{R}R^{-1}$ again but the constant
$C_1=0$. Intriguingly, the Copernican principle is automatically
satisfied.

Unfortunately, it is very difficult to reproduce acceleration
equation \eqref{Friedmann1} from gravitational equations
\eqref{Einstein-FLRW}. From \eqref{Einstein-FLRW} we obtain
%\begin{equation}
%\xi^{\mu}\partial_{\mu}=\partial_{t}-\frac{\ddot{R}r}{\dot{R}}\partial_{r}.
%\end{equation}
\begin{equation}
-\frac{3\ddot{R}}{R}f-\partial_{t}^2f+\frac{\ddot{R}r}{\dot{R}}\partial_{t}\partial_{r}f-\frac{\ddot{R}r}{R}\partial_{r}f=4\pi G(\rho+3p).
\end{equation}
This equation can be integrated to give \eqref{Friedmann1} if
redundant terms vanish,
\begin{equation}\label{consistency}
-\partial_{t}^2f+\frac{\ddot{R}r}{\dot{R}}\partial_{t}\partial_{r}f=0.
\end{equation}
For a general function of $f$, the condition is not always satisfied
and hence we cannot recover Eq. \eqref{Friedmann1}. This discrepancy
is not attributed to any error in our calculation. It stems from a
technical trick in Sec. \ref{subsect-rel}: we began with a static
background which has a time-like Killing vector
\cite{Verlinde:2010hp}. For the FLRW spacetime, this is possible
only if metric \eqref{FLRW} reduces to the de Sitter or Minkowski
spacetime \cite{YuWang}. Indeed, it is checkable that for a general
function $f(a,A)$, condition \eqref{consistency} is assured if
$R=\exp(Ht)$ with a constant $H$. In other words, Eq.
\eqref{Friedmann1} and Eqs. \eqref{Einstein-FLRW} are consistent in
static cases, though the inconsistency persists in other cases.

The issue discussed above present a new challenge to modified models
of entropic gravity. So far we do not have a perfect solution to
this challenge. At this point we mention several possibilities.
First, possibly the FLRW metric is not an exact solution of modified
entropic gravity models but needs adjustment. The second possible
attitude is insisting on \eqref{Friedmann1} and taking
\eqref{consistency} as a consistency condition for allowable models.
Note that this condition is met by Einstein gravity ($f=1$). The
third possible way to alleviate the contradiction is tuning the
definition of red-shifted ``temperature'' for the FLRW universe in
modified entropic gravity,
\begin{equation}
T\propto-\ddot{R}r+\frac{R}{\dot{R}r^2f}\int_{\mathcal{V}}\left[\partial_{t}(\dot{R}r)\partial_{r}\partial_{t}f-\partial_{r}(\dot{R}r)\partial_{t}^2f\right]r^2dr,
\end{equation}
giving rise to a correction term to \eqref{Friedmann1}. The fourth
but very difficult solution is deriving gravitational equations
without assuming a static background, then probably we will get
field equations different from \eqref{Einstein}. Before arriving at
the final answer, it is still too early to claim which possibility
is more promising or whether there are other possibilities.

%\begin{equation}
%\partial_{r}=-\ddot{R}\partial_{a}+8\pi R^2r\partial_{A},~~~\partial_{t}=-\dddot{R}r\partial_{a}+8\pi R\dot{R}r^2\partial_{A}.
%\end{equation}
%\begin{equation}
%\alpha\left(-\frac{3\ddot{R}}{R}f-\partial_{t}^2f\right)+\beta r\left(\partial_{t}\partial_{r}f-\frac{\dot{R}}{R}\partial_{r}f\right)=4\pi G(\rho+3p).
%\end{equation}
%\begin{equation}
%\frac{\beta}{\alpha}=\frac{\dot{R}}{R}=\frac{\ddot{R}}{\dot{R}}=\frac{\dddot{R}}{\ddot{R}}.
%\end{equation}
%\begin{eqnarray}
%\nonumber \nabla^2f&=&-\frac{1}{R^3}\partial_{t}(R^3\partial_{t}f)+\frac{1}{R^2r^2}\partial_{r}(r^2\partial_{r}f)\\
%&=&-\partial_{t}^2f-\frac{3\dot{R}}{R}\partial_{t}f+\frac{3}{R^2r}\partial_{r}f.
%\end{eqnarray}
In accordance with gravitational equations \eqref{Einstein}, we can
also study condition \eqref{emcons} of covariant energy-momentum
conservation. Assuming the FLRW metric \eqref{FLRW}, this condition
becomes
\begin{equation}
6\left(\frac{\ddot{R}}{R}+\frac{\dot{R}^2}{R^2}\right)\nabla_{\mu}f+\nabla_{\mu}\nabla^2f=0,
\end{equation}
or explicitly
\begin{eqnarray}
\nonumber \left(\frac{5\ddot{R}}{R}+\frac{3\dot{R}^2}{R^2}\right)\partial_{r}f&=&0,\\
\left(\frac{3\ddot{R}}{R}+\frac{9\dot{R}^2}{R^2}\right)\partial_{t}f-\partial_{t}^3f-\frac{3\dot{R}}{R}\partial_{t}^2f-\frac{3\dot{R}}{R^3r}\partial_{r}f&=&0.
\end{eqnarray}
Putting \eqref{f2} into these equations, we are led to a severe
constraint. The constraint rules out nearly all models of form
\eqref{EaAf} other than Einstein gravity.

\section{Discussion}\label{sect-disc}
In this paper, we unified a number of modified entropic gravity
models in the literature to a general form \eqref{EaAf}. The
corresponding gravitational equations are demonstrated to be
\eqref{Einstein}, whose static weak field limit is consistent with a
straightforward calculation. If the modified model deviates
significantly from Einstein gravity or Newtonian gravity, the static
weak field equation \eqref{Gaussd} provides an arena to test them
against observations.

In these models, the energy-momentum tensor is covariantly conserved
if and only if condition \eqref{emcons} is met. The condition can be
regarded as a constraint on viable models, or reverse the logic, a
constraint on suitable metric solutions for modified entropic
gravity.

Assuming the FLRW metric, we derived cosmological equations in two
independent approaches. To our surprise, a discrepancy exists unless
condition \eqref{consistency} is satisfied. There are several
possible ways to reconcile this discrepancy. One way is taking
\eqref{consistency} as a consistency condition of admissible models,
then Einstein gravity ($f=1$) wins out.

As indicated by our results, the modified entropic gravity models of
form \eqref{EaAf}, if not killed, should live in a very narrow room
to assure the energy-momentum conservation and to accommodate a
homogeneous isotropic universe.

\begin{acknowledgments}
This work is supported by the NSFC grant No.11105053.
\end{acknowledgments}

\appendix

\section{Useful formulae}\label{app-form}
%in apparent contradiction to, the discrepancy arises, $R(t)=e^{Ht}$
In this appendix, we gather some formulae for spatially flat
FLRW spacetime \eqref{FLRW}. These formulae are useful in Sec. \ref{sect-cosmo}.
%\begin{equation}\label{FLRW}
%ds^2=-dt^2+R^2\left[\frac{dr^2}{1-kr^2}+r^2(d\theta^2+\sin^2\theta d\varphi^2)\right].
%\end{equation}

In our notations, the nonvanishing Christoffel connections are
\begin{eqnarray}
\nonumber &&\Gamma^{t}_{rr}=R\dot{R},~~~\Gamma^{t}_{\theta\theta}=R\dot{R}r^2,~~~\Gamma^{t}_{\varphi\varphi}=R\dot{R}r^2\sin^2\theta,\\
\nonumber &&\Gamma^{r}_{\theta\theta}=-r,~~~\Gamma^{r}_{\varphi\varphi}=-r\sin^2\theta,~~~\Gamma^{\theta}_{\varphi\varphi}=-\sin\theta\cos\theta,\\
&&\Gamma^{r}_{tr}=\Gamma^{\theta}_{t\theta}=\Gamma^{\varphi}_{t\varphi}=\frac{\dot{R}}{R},~~~\Gamma^{\theta}_{r\theta}=\Gamma^{\varphi}_{r\varphi}=\frac{1}{r},~~~\Gamma^{\varphi}_{\theta\varphi}=\cot\theta,
\end{eqnarray}
which lead to the nonzero components of the Ricci tensor
\begin{eqnarray}
\nonumber &&\mathcal{R}_{tt}=-\frac{3\ddot{R}}{R},~~~\mathcal{R}_{rr}=R\ddot{R}+2\dot{R}^2,\\
&&\mathcal{R}_{\theta\theta}=\left(R\ddot{R}+2\dot{R}^2\right)r^2,~~~\mathcal{R}_{\varphi\varphi}=\left(R\ddot{R}+2\dot{R}^2\right)r^2\sin^2\theta
\end{eqnarray}
and the Ricci scalar
\begin{equation}
\mathcal{R}=6\left(\frac{\ddot{R}}{R}+\frac{\dot{R}^2}{R^2}\right).
\end{equation}

When $H=\dot{R}/R$ is restricted to be constant, the FLRW metric
reduces to the de Sitter spacetime. This spacetime has a timelike
Killing vector
\begin{equation}
\xi^{\mu}\partial_{\mu}=\partial_{t}-Hr\partial_{r},
\end{equation}
giving rise to $e^{2\phi}=1-H^2R^2r^2$. The other useful formulae
for the de Sitter spacetime are
\begin{eqnarray}
%\nonumber &&\mathcal{N}^{\mu}\partial_{\mu}=(1-H^2R^2r^2)^{-1/2}\left(-HRr\partial_{t}+\frac{1}{R}\partial_{r}\right),\\
\nonumber &&\mathcal{N}^{\mu}\partial_{\mu}=e^{-\phi}\left(-HRr\partial_{t}+\frac{1}{R}\partial_{r}\right),\\
&&u^{\mu}\partial_{\mu}=n^{\mu}\partial_{\mu}=\partial_{t}.
\end{eqnarray}
It is trivial to check
$a=e^{\phi}\mathcal{N}^{\mu}\nabla_{\mu}\phi=-H^2Rr$.


\begin{thebibliography}{99}
\bibitem{Jacobson:1995ab}
  T.~Jacobson,
  %``Thermodynamics of space-time: The Einstein equation of state,''
  Phys.\ Rev.\ Lett.\  {\bf 75}, 1260 (1995)
  [arXiv:gr-qc/9504004].
  %%CITATION = PRLTA,75,1260;%%

\bibitem{Verlinde:2010hp}
  E.~P.~Verlinde,
  %``On the Origin of Gravity and the Laws of Newton,''
  JHEP {\bf 1104}, 029 (2011)
  [arXiv:1001.0785 [hep-th]].
  %%CITATION = ARXIV:1001.0785;%%

\bibitem{Unruh:1976db}
  W.~G.~Unruh,
  %``Notes on black hole evaporation,''
  Phys.\ Rev.\ D {\bf 14}, 870 (1976).
  %%CITATION = PHRVA,D14,870;%%

\bibitem{Padmanabhan:2009kr}
  T.~Padmanabhan,
  %``Equipartition of energy in the horizon degrees of freedom and the emergence
  %of gravity,''
  arXiv:0912.3165 [gr-qc].
  %%CITATION = ARXIV:0912.3165;%%

\bibitem{Zhao:2010qw}
  L.~Zhao,
  %``Hidden symmetries for thermodynamics and emergence of relativity,''
  Commun.\ Theor.\ Phys.\  {\bf 54}, 641 (2010)
  [arXiv:1002.0488 [hep-th]].
  %%CITATION = ARXIV:1002.0488;%%

\bibitem{Tian:2010uy}
  Y.~Tian and X.~-N.~Wu,
  %``Thermodynamics of Black Holes from Equipartition of Energy and Holography,''
  Phys.\ Rev.\ D {\bf 81}, 104013 (2010)
  [arXiv:1002.1275 [hep-th]].
  %%CITATION = ARXIV:1002.1275;%%

\bibitem{Liu:2010na}
  Y.~-X.~Liu, Y.~-Q.~Wang and S.~-W.~Wei,
  %``A Note on Temperature and Energy of 4-dimensional Black Holes from Entropic Force,''
  Class.\ Quant.\ Grav.\  {\bf 27}, 185002 (2010)
  [arXiv:1002.1062 [hep-th]].
  %%CITATION = ARXIV:1002.1062;%%

\bibitem{KowalskiGlikman:2010ms}
  J.~Kowalski-Glikman,
  %``A note on gravity, entropy, and BF topological field theory,''
  Phys.\ Rev.\ D {\bf 81}, 084038 (2010)
  [arXiv:1002.1035 [hep-th]].
  %%CITATION = ARXIV:1002.1035;%%

\bibitem{Myung:2010jv}
  Y.~S.~Myung,
  %``Entropic Force in the Presence of Black Hole,''
  arXiv:1002.0871 [hep-th].
  %%CITATION = ARXIV:1002.0871;%%

\bibitem{Konoplya:2010ak}
  R.~A.~Konoplya,
  %``Entropic force, holography and thermodynamics for static space-times,''
  Eur.\ Phys.\ J.\ C {\bf 69}, 555 (2010)
  [arXiv:1002.2818 [hep-th]].
  %%CITATION = ARXIV:1002.2818;%%

\bibitem{Gao:2010yy}
  S.~Gao,
  %``Is Gravity an Entropic Force?,''
  Entropy {\bf 13}, 936 (2011)
  [arXiv:1002.2668 [physics.gen-ph]].
  %%CITATION = ARXIV:1002.2668;%%

\bibitem{Cai:2010zw}
  Y.~-F.~Cai, J.~Liu and H.~Li,
  %``Entropic cosmology: a unified model of inflation and late-time acceleration,''
  Phys.\ Lett.\ B {\bf 690}, 213 (2010)
  [arXiv:1003.4526 [astro-ph.CO]].
  %%CITATION = ARXIV:1003.4526;%%

\bibitem{Ho:2010ca}
  C.~M.~Ho, D.~Minic and Y.~J.~Ng,
  %``Cold Dark Matter with MOND Scaling,''
  Phys.\ Lett.\ B {\bf 693}, 567 (2010)
  [arXiv:1005.3537 [hep-th]].
  %%CITATION = ARXIV:1005.3537;%%

\bibitem{Pikhitsa:2010nd}
  P.~V.~Pikhitsa,
  %``MOND reveals the thermodynamics of gravity,''
  arXiv:1010.0318 [astro-ph.CO].
  %%CITATION = ARXIV:1010.0318;%%

\bibitem{Klinkhamer:2011un}
  F.~R.~Klinkhamer and M.~Kopp,
  %``Entropic gravity, minimum temperature, and modified Newtonian dynamics,''
  Mod.\ Phys.\ Lett.\ A {\bf 26}, 2783 (2011)
  [arXiv:1104.2022 [hep-th]].
  %%CITATION = ARXIV:1104.2022;%%

\bibitem{Zhang:2010hi}
  Y.~Zhang, Y.~Gong and Z.~-H.~Zhu,
  %``Modified gravity emerging from thermodynamics and holographic principle,''
  Int.\ J.\ Mod.\ Phys.\ D {\bf 20}, 1505 (2011)
  [arXiv:1001.4677 [hep-th]].
  %%CITATION = ARXIV:1001.4677;%%

\bibitem{Wei:2010wwa}
  S.~-W.~Wei, Y.~-X.~Liu and Y.~-Q.~Wang,
  %``A note on Friedmann equation of FRW universe in deformed Horava-Lifshitz gravity from entropic force,''
  Commun.\ Theor.\ Phys.\  {\bf 56}, 455 (2011)
  [arXiv:1001.5238 [hep-th]].
  %%CITATION = ARXIV:1001.5238;%%

\bibitem{Modesto:2010rm}
  L.~Modesto and A.~Randono,
  %``Entropic Corrections to Newton's Law,''
  arXiv:1003.1998 [hep-th].
  %%CITATION = ARXIV:1003.1998;%%

\bibitem{Myung:2010kd}
  Y.~S.~Myung,
  %``Does entropic force always imply the Newtonian force law?,''
  Eur.\ Phys.\ J.\ C {\bf 71}, 1549 (2011)
  [arXiv:1003.5037 [hep-th]].
  %%CITATION = ARXIV:1003.5037;%%

\bibitem{Sheykhi:2010wm}
  A.~Sheykhi,
  %``Entropic Corrections to Friedmann Equations,''
  Phys.\ Rev.\ D {\bf 81}, 104011 (2010)
  [arXiv:1004.0627 [gr-qc]].
  %%CITATION = ARXIV:1004.0627;%%

\bibitem{Pan:2010eu}
  Q.~Pan and B.~Wang,
  %``Influence on the entropic force by the virtual degree of freedom on the holographic screen,''
  Phys.\ Lett.\ B {\bf 694}, 456 (2011)
  [arXiv:1004.2954 [hep-th]].
  %%CITATION = ARXIV:1004.2954;%%

\bibitem{Setare:2010ct}
  M.~R.~Setare, D.~Momeni and R.~Myrzakulov,
  %``Entropic corrections to Newton's law,''
  Phys.\ Scripta {\bf 85}, 065007 (2012)
  [arXiv:1004.2794 [physics.gen-ph]].
  %%CITATION = ARXIV:1004.2794;%%

\bibitem{Liu:2010zn}
  B.~Liu, Y.~-C.~Dai, X.~-R.~Hu and J.~-B.~Deng,
  %``The Friedmann equation in modified entropy-area relation from entropy force,''
  Mod.\ Phys.\ Lett.\ A {\bf 26}, 489 (2011)
  [arXiv:1010.3429 [hep-th]].
  %%CITATION = ARXIV:1010.3429;%%

\bibitem{Sheykhi:2010yq}
  A.~Sheykhi and S.~H.~Hendi,
  %``Power-Law Entropic Corrections to Newton's Law and Friedmann Equations,''
  Phys.\ Rev.\ D {\bf 84}, 044023 (2011)
  [arXiv:1011.0676 [hep-th]].
  %%CITATION = ARXIV:1011.0676;%%

\bibitem{Hendi:2010xr}
  S.~H.~Hendi and A.~Sheykhi,
  %``Entropic Corrections to Einstein Equations,''
  Phys.\ Rev.\ D {\bf 83}, 084012 (2011)
  [arXiv:1012.0381 [hep-th]].
  %%CITATION = ARXIV:1012.0381;%%

\bibitem{ChangYoung:2011rb}
  E.~Chang-Young, K.~Kimm and D.~Lee,
  %``Brans-Dicke Gravity as an Entropic Phenomenon,''
  arXiv:1105.5905 [hep-th].
  %%CITATION = ARXIV:1105.5905;%%

\bibitem{Gao:2010fw}
  C.~Gao,
  %``Modified Entropic Force,''
  Phys.\ Rev.\ D {\bf 81}, 087306 (2010)
  [arXiv:1001.4585 [hep-th]].
  %%CITATION = ARXIV:1001.4585;%%

\bibitem{Li:2010si}
  X.~Li and Z.~Chang,
  %``Debye entropic force and modified Newtonian dynamics,''
  Commun.\ Theor.\ Phys.\  {\bf 55}, 733 (2011)
  [arXiv:1005.1169 [hep-th]].
  %%CITATION = ARXIV:1005.1169;%%

\bibitem{Wei:2010am}
  H.~Wei,
  %``Cosmological Constraints on the Modified Entropic Force Model,''
  Phys.\ Lett.\ B {\bf 692}, 167 (2010)
  [arXiv:1005.1445 [gr-qc]].
  %%CITATION = ARXIV:1005.1445;%%

\bibitem{Chang:2010be}
  Z.~Chang, M.~-H.~Li and X.~Li,
  %``Unification of Dark Matter and Dark Energy in a Modified Entropic Force Model,''
  Commun.\ Theor.\ Phys.\  {\bf 56}, 184 (2011)
  [arXiv:1009.1506 [gr-qc]].
  %%CITATION = ARXIV:1009.1506;%%

\bibitem{Kiselev:2010xi}
  V.~V.~Kiselev and S.~A.~Timofeev,
  %``The Holographic screen at low temperatures,''
  Mod.\ Phys.\ Lett.\ A {\bf 26}, 109 (2011)
  [arXiv:1009.1301 [hep-th]].
  %%CITATION = ARXIV:1009.1301;%%

\bibitem{Neto:2010ds}
  J.~A.~Neto,
  %``Nonhomogeneous Cooling, Entropic Gravity and MOND Theory,''
  Int.\ J.\ Theor.\ Phys.\  {\bf 50}, 3552 (2011)
  [arXiv:1009.4944 [hep-th]].
  %%CITATION = ARXIV:1009.4944;%%

\bibitem{Abreu:2012ax}
  E.~M.~C.~Abreu, J.~A.~Neto, A.~C.~R.~Mendes and W.~Oliveira,
  %``Nonextensive Friedmann equations and new bounds for Tsallis parameter through noncommutative entropic gravity,''
  arXiv:1204.2005 [hep-th].
  %%CITATION = ARXIV:1204.2005;%%

\bibitem{SheykhiSarab}
  A. Sheykhi, K. R. Sarab,
  %``Einstein Equations and MOND Theory from Debye Entropic Gravity,''
  arXiv:1206.1030 [physics.gen-ph].
  %%CITATION = ARXIV:1206.1030;%%

\bibitem{Wang:2012pk}
  T.~Wang, Y.~-W.~Wu and J.~Zhao,
  %``A quantum oscillator model for entropic gravity,''
  arXiv:1207.6041 [hep-th].
  %%CITATION = ARXIV:1207.6041;%%

\bibitem{Liu:2010wp}
  B.~Liu, Y.~-C.~Dai, X.~-R.~Hu and J.~-B.~Deng,
  %``The Friedmann equations in deformed Ho\v{r}ava-Lifshitz gravity and Debye model,''
  Int.\ J.\ Mod.\ Phys.\ D {\bf 20}, 497 (2011)
  [arXiv:1007.2985 [hep-th]].
  %%CITATION = ARXIV:1007.2985;%%

\bibitem{Vancea:2010vf}
  I.~V.~Vancea and M.~A.~Santos,
  %``Entropic Law of Force, Emergent Gravity and the Uncertainty Principle,''
  Mod.\ Phys.\ Lett.\ A {\bf 27}, 1250012 (2012)
  [arXiv:1002.2454 [hep-th]].
  %%CITATION = ARXIV:1002.2454;%%

\bibitem{Ghosh:2010hz}
  S.~Ghosh,
  %``Planck Scale Effect in the Entropic Force Law,''
  arXiv:1003.0285 [hep-th].
  %%CITATION = ARXIV:1003.0285;%%

\bibitem{ShalytMargolin:2012zh}
  A.~E.~Shalyt-Margolin,
  %``Probable Entropic Nature of Gravity in Ultraviolet and Infrared Limits. I.Ultraviolet Case,''
  arXiv:1205.6988 [hep-th].
  %%CITATION = ARXIV:1205.6988;%%

\bibitem{Jackiw:2003pm}
  R.~Jackiw and S.~Y.~Pi,
  %``Chern-Simons modification of general relativity,''
  Phys.\ Rev.\ D {\bf 68}, 104012 (2003)
  [gr-qc/0308071].
  %%CITATION = GR-QC/0308071;%%

\bibitem{Shu:2010nv}
  F.~-W.~Shu and Y.~Gong,
  %``Equipartition of energy and the first law of thermodynamics at the apparent horizon,''
  Int.\ J.\ Mod.\ Phys.\ D {\bf 20}, 553 (2011)
  [arXiv:1001.3237 [gr-qc]].
  %%CITATION = ARXIV:1001.3237;%%

\bibitem{Cai:2010hk}
  R.~G.~Cai, L.~M.~Cao and N.~Ohta,
  %``Friedmann Equations from Entropic Force,''
  Phys.\ Rev.\  D {\bf 81}, 061501 (2010)
  [arXiv:1001.3470 [hep-th]].
  %%CITATION = PHRVA,D81,061501;%%

\bibitem{Padmanabhan:2003pk}
  T.~Padmanabhan,
  %``Gravitational entropy of static space-times and microscopic density of states,''
  Class.\ Quant.\ Grav.\  {\bf 21}, 4485 (2004)
  [gr-qc/0308070].
  %%CITATION = GR-QC/0308070;%%

\bibitem{YuWang}
  R.~Y.~Yu and T.~Wang,
  ``Studying cosmological apparent horizon with quasi-static coordinates,''
  to appear in Pramana {\bf 80},  (2013).
  %%CITATION = PHRVA,D81,104045;%%

\end{thebibliography}
\end{document}